\documentclass[acmsmall,screen]{acmart}

\setcopyright{none}             
\renewcommand\footnotetextcopyrightpermission[1]{} 

\author{Lezhi Ma}
\affiliation{
  \institution{Nanjing University}
  \city{Nanjing}
  \country{China}
}
\email{lezhima@hotmail.com}

\author{Han Wang}
\affiliation{
  \institution{Nanjing University}
  \city{Nanjing}
  \country{China}
}
\email{wh66638@outlook.com}

\author{Shangqing Liu}
\affiliation{
  \institution{Nanjing University}
  \city{Nanjing}
  \country{China}
}
\email{shangqingliu@nju.edu.cn}

\author{Jiawan Wang}
\affiliation{
  \institution{Nanjing University}
  \city{Nanjing}
  \country{China}
}
\email{wjwnju@nju.edu.cn}

\author{Lei Bu}
\affiliation{
  \institution{Nanjing University}
  \city{Nanjing}
  \country{China}
}
\email{bulei@nju.edu.cn}

\usepackage{hyperref}
\hypersetup{hidelinks}
\usepackage{graphicx} 
\usepackage{svg} 
\usepackage{amsfonts}
\usepackage{amsmath}
\usepackage{enumerate}
\usepackage{enumitem}
\usepackage{xspace}
\usepackage{subcaption}
\usepackage{booktabs}
\usepackage{multirow}
\usepackage{multicol}
\usepackage[most]{tcolorbox}
\usepackage{listings}
\usepackage{algorithm}
\usepackage{algpseudocode}

\newtcolorbox{promptresponsebox}[1][]{
  colback=white,
  colframe=black,
  fonttitle=\bfseries,
  coltitle=black,
  colbacktitle=gray!20,
  title={Prompt-Response Example},
  boxrule=0.5pt,
  top=1pt,
  bottom=1pt,
  left=4pt,
  right=4pt,
  toptitle=0.5pt,
  bottomtitle=0.5pt,
  #1
}

\newcommand{\tool}[0]{\textit{SpecAgent}\xspace}

\title{\tool: Empowering Program Verification with Agentic Synthesis of Formal Program Specifications}

\begin{abstract}
Formal specifications are essential for deductive program verification, as they provide semantic abstractions that enable compositional verification on complex software systems. The difficulties in manually constructing such specifications gave birth to automated specification synthesis, and recent advances in relevant approaches based on large language models (LLMs) have shown promising capabilities. Nevertheless, existing approaches often follow a forward-only synthesis workflow and rely on localized repair strategies, which limit their effectiveness on real-world programs with complex dependencies among functions and loops. In such scenarios, verification failures may originate from previously generated specifications rather than the immediate failure locations, leading to cascading failures that cannot be effectively addressed by local refinement.

In this paper, we present \tool, an agentic framework for synthesizing high-quality ACSL specifications for real-world C programs. \tool integrates four complementary components into a unified workflow: it analyzes program structures and dependencies to plan specification targets, leverages retrieval-augmented generation with a specification knowledge base to provide LLMs with relevant specification patterns and program context, employs agentic repairing to diagnose specification defects and revisit previously generated specifications when verification failures are traced to their dependencies, and applies an agentic critique mechanism to assess and refine semantic strength beyond proof success.
We extensively evaluate \tool on both specification synthesis and program verification tasks. On a benchmark consisting of 50 programs from 14 real-world repositories, \tool with DeepSeek-V4 achieves 96.82\% precision and 88.95\% recall in synthesizing correct yet strong specifications, substantially outperforming all baselines including Codex equipped with GPT-5.6. We further evaluate \tool on 811 real-world verification targets, where \tool successfully discharges 604 targets, surpassing existing baselines. The results show that \tool can effectively synthesize high-quality specifications and facilitate verification of real-world software.
\end{abstract}

\begin{document}

\maketitle

\section{Introduction}

Program verification~\cite{hoare1969axiomatic} is an essential technique for software reliability assurance, which can provide rigorous guarantees by proving consistency between the target program and given properties (including functional correctness and runtime safety constraints), complementing conventional techniques such as software testing~\cite{howden1991program}. However, automatically verifying complex programs against high-level properties is fundamentally challenging, as the general program verification problem is undecidable~\cite{mathur2019decidable} due to the potential unbounded behaviors produced by intricate program structures such as loops and function calls. One typical approach to achieving deductive verification is based on \emph{auxiliary specifications}~\cite{frade2011verification}, which are expressed in parsable, unambiguous formal languages and can abstract the behaviors of individual program structures. Specifically, loop invariants summarize the properties preserved across loop iterations, while function contracts (typically including preconditions and postconditions) characterize the behavioral guarantees of function calls. These specifications provide the semantic abstractions needed by the verifier to construct local proof obligations and facilitate compositional proof construction, thereby enabling the verification of high-level target properties of the overall program~\cite{frade2011verification,henzinger1998you}.

However, manually crafting formal program specifications remains labor-intensive and expertise-demanding~\cite{schiller2012reducing,snook2001exploring}, particularly for real-world software with complicated control flows and intricate data dependencies. This challenge has motivated substantial research into automated specification synthesis. Early approaches in this direction typically rely on heuristic rules~\cite{flanagan2001houdini,ernst2007daikon,molina2022fuzzing} or machine learning techniques~\cite{li2017automatic,si2018learning,ryan2019cln2inv}. Recently, large language models (LLMs)~\cite{openai2023gpt,openai2025gpt,grattafiori2024llama,liu2024deepseek,guo2025deepseek,xu2026deepseek} have demonstrated promising capabilities in program comprehension and code summarization, bringing new light on automated program specification synthesis. A series of LLM-based approaches~\cite{wen2024enchanting,ma2024specgen,pei2023can,cao2025informal,wang2025tale} have been explored to generate function contracts and loop invariants, often in conjunction with formal verification tools to validate and refine the generated specifications.

Despite this progress, synthesizing high-quality specifications for complex real-world programs remains challenging due to fundamental limitations of existing approaches. First, they typically follow a \emph{forward-only} workflow, in which a specification is considered finalized once successfully generated and is not revisited in subsequent stages. Second, their repair process is predominantly \emph{localized}: when verification fails, they locate and repair errors solely based on the locations indicated by the verifier, without considering specifications at other program points. These design choices become problematic in complex real-world programs, where intricate dependencies widely exist among program structures including loops and function calls, and consequently among the specifications that abstract their behaviors. Due to these dependencies, verification failure in downstream specifications may stem from semantic correctness/strength issues in other upstream specifications on which they depend. Such issues can propagate along dependency chains, triggering \emph{cascading verification failures}. However, the forward-only workflow prevents existing approaches from revisiting upstream specifications, while localized repair confines their efforts to the immediate failure site. Consequently, they may repeatedly attempt ineffective local repairs without addressing the root cause, compromising the correctness and strength of the synthesized specifications and ultimately limiting their ability to support the verification of target properties.

Addressing this limitation presents several challenges. First, specification synthesis must account for dependencies induced by program structure. Real-world programs involve intricate relationships among functions, loops, and data structures, making it difficult to identify the specification synthesis order and the underlying specifications responsible for verification failures. Second, verification feedback does not directly identify the root causes of specification defects. A failed proof may result from an incorrect or insufficient specification, a runtime safety violation, or limitations of the underlying verification process. Distinguishing these causes and determining the appropriate scope of refinement require reasoning over both verification evidence and program context. Third, proof success alone is insufficient to establish specification quality. Weak or vacuous specifications may be readily verified while failing to capture meaningful program behavior or provide useful guarantees to callers. Therefore, an effective synthesis approach must not only produce verifiable specifications but also assess their semantic adequacy and refine them when necessary.


These challenges call for a shift from localized specification generation and repair toward \emph{agentic specification synthesis}, where specifications are generated and refined through agentic dependency-aware decision-making. Instead of treating generated specifications as immutable artifacts, such a process can trace root causes across specification dependencies, revisit and improve defective specifications, and ensure that the resulting specifications provide meaningful behavioral abstractions beyond merely passing verification. To realize this vision, we propose \tool, an agentic framework for synthesizing formal specifications in ACSL for real-world C programs. \tool first employs dependency-aware planning to identify specification targets and organize their synthesis according to program structure and verification objectives. It then leverages agentic generation, augmented with relevant specification knowledge and program context, to construct auxiliary specifications, which are subsequently checked by formal verification tools, guiding the system in diagnosing failures and determining appropriate repair actions. Unlike workflows restricted to localized repair, \tool utilizes agentic repairing that can trace verification issues across program dependencies and revisit previously generated specifications when they are identified as potential root causes. In addition, a specification critique mechanism evaluates the semantic strength of verified specifications, helping identify cases in which proof success does not translate into sufficiently informative behavioral guarantees. By integrating these capabilities into a unified workflow, \tool aims to improve both the correctness and quality of specifications while reducing the manual effort required for verifying real-world software.

We evaluate SpecAgent through extensive experiments from two complementary perspectives: quality of synthesized specifications and ability to support program verification. For the specification synthesis task, we conduct experiments on a benchmark comprising 50 sizable programs from 14 real-world C repositories, where \tool achieves 96.82\% precision and 88.95\% recall, substantially outperforming existing approaches (including Codex with \texttt{gpt-5.6-sol}) in generating correct and sufficiently complete specifications. We further evaluate the ability of \tool-generated specifications to support real-world verification tasks on 811 verification targets from two established benchmarks, covering both runtime-safety and reachability-oriented verification tasks. \tool successfully discharges 604 out of 811 verification targets, substantially surpassing the best-performing existing baseline (yielding 523). These results provide empirical evidence that the specifications synthesized by SpecAgent are not only semantically accurate but also practically effective in facilitating the verification of high-level target properties.

To conclude, the main contributions of this paper include:
\begin{itemize}[leftmargin=*]
    \item \tool, an agentic framework integrating dependency-aware planning, retrieval-augmented specification generation, formal verification, and agentic repair and refinement to synthesize ACSL specifications for real-world C program verification tasks.
    \item An agentic specification repair and refinement workflow that leverages agentic decisions to diagnose specification defects (in terms of both semantic correctness and strength), enabling effective revisiting and refinement of previously generated specifications. 
    \item Extensive experiments on 14 repositories and 811 real-world verification targets, comparing \tool with 5 state-of-the-art baselines to investigate its effectiveness, component contributions, and practical applicability.
\end{itemize}
\section{Background and Motivation}
\label{sec:background}

\subsection{Program Verification and Formal Specification Synthesis}

Program verification~\cite{hoare1969axiomatic} aims to establish whether a program satisfies desired properties, such as functional correctness and runtime safety constraints, by formally proving the consistency between program behaviors and given specifications. However, directly verifying high-level properties of complex programs is challenging due to the unbounded behaviors introduced by intricate control-flow structures, such as loops and function calls~\cite{mathur2019decidable}. To enable effective verification, deductive verification commonly relies on \emph{auxiliary specifications}~\cite{frade2011verification}, which describe the behaviors of individual program structures in precise and unambiguous formal languages. These specifications serve as semantic abstractions that allow verifiers to encode program behaviors compositionally. For example, function contracts summarize the assumptions and guarantees of function invocations, while loop invariants characterize properties preserved across loop iterations. By providing such abstractions, auxiliary specifications enable verifiers to construct local verification obligations and combine them to establish the proof for high-level properties of the overall program.

In this work, we focus on the ANSI/ISO C Specification Language (ACSL)~\cite{baudin2008acsl}, a formal specification language for annotating C programs. ACSL provides various constructs for describing program behaviors, among which function contracts and loop invariants are particularly important for deductive verification. Specifically, \texttt{requires} statements define preconditions that must hold before function execution, \texttt{ensures} statements specify postconditions guaranteed after execution, and \texttt{loop invariant} statements describe properties maintained throughout loop iterations. These annotations provide formal behavioral summaries of program structures, which can be leveraged by verification frameworks such as Frama-C/WP~\cite{patrick2025wp} to generate verification conditions and discharge them using automated theorem proving and SMT solving~\cite{de2008z3,barrett2011cvc4,conchon2018alt}.

Despite their importance, constructing effective auxiliary specifications remains challenging. Writing formal specifications requires developers to not only understand the intended behaviors of programs but also identify appropriate abstractions and express them in a rigorous formal language~\cite{schiller2012reducing}. For real-world software with entangled data dependencies and numerous interacting components, manually designing sufficiently precise specifications can be labor-intensive and error-prone. This has motivated extensive research on automated specification synthesis. Traditional specification synthesis approaches range from rule-based techniques~\cite{flanagan2001houdini,ernst2007daikon,molina2022fuzzing} to learning-based methods~\cite{li2017automatic,si2018learning,ryan2019cln2inv}. More recently, the rapid development of Large Language Models (LLMs)~\cite{openai2023gpt,achiam2023gpt,openai2025gpt,liu2024deepseek,guo2025deepseek,grattafiori2024llama} has brought new opportunities for program specification generation by leveraging their strong capabilities in code understanding and summarization, giving rise to a series of LLM-based approaches~\cite{wen2024enchanting,ma2024specgen,wang2025tale,ma2026specsyn}.

\subsection{Motivation}

Real-world programs rarely consist of independent components. Functions interact through calls and shared data, data structures are constrained by the operations that manipulate them, and loops and their surrounding code jointly determine program states and invariants. Consequently, the specifications of individual program components are often semantically interdependent: a specification for one component often relies on preconditions, postconditions, invariants, or behavioral properties established by another. Effective specification synthesis therefore needs to account for such dependencies and propagate sufficient semantic information across them.

\begin{figure}[!t]
    \centering
    \includegraphics[width=0.93\linewidth]{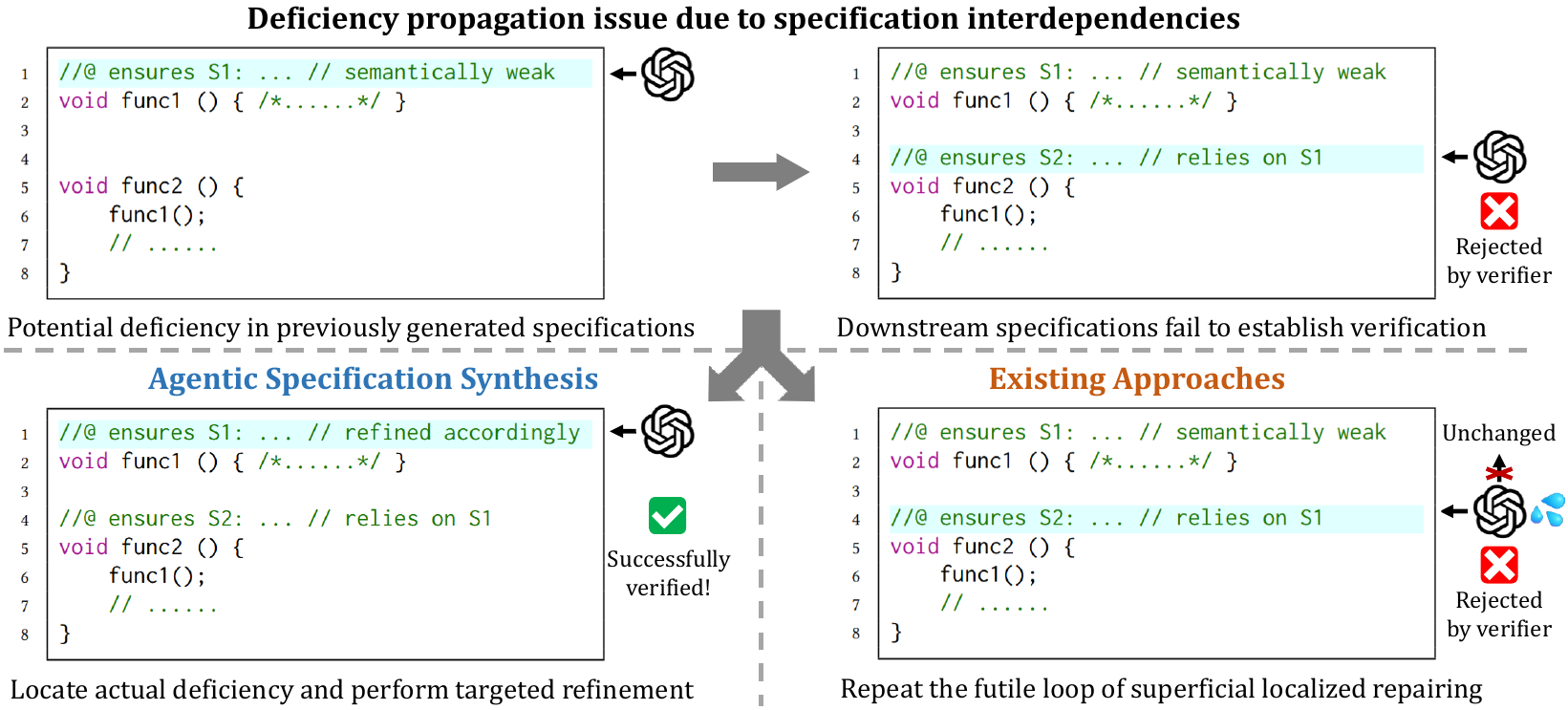}
    \vspace{-2mm}
    \caption{Illustration of specification deficiency propagation. Since \texttt{func2} invokes \texttt{func1}, its postcondition \texttt{S2} also depends on the postcondition \texttt{S1} of \texttt{func1}, and they are typically processed sequentially. Semantic deficiencies in \texttt{S1} will result in verification failures of \texttt{S2}, where existing approaches repeatedly perform ineffective localized repairs on \texttt{S2} due to their forward-only practice. In contrast, agentic synthesis can reason about the root cause to effectively revisit and refine the responsible upstream specification \texttt{S1}.}
    \vspace{-4mm}
    \label{fig:motivation}
\end{figure}

Existing specification synthesis approaches, such as Preguss~\cite{wang2025tale} and SpecSyn~\cite{ma2026specsyn}, have been aware of the phenomenon and choose to generate specifications incrementally according to the dependency relations among program components. Nevertheless, they largely follow a \textit{forward-only} process: once a specification has been generated and satisfies the verification checks available at that stage, it is typically treated as finalized and will never be revisited. This creates a potential problem when an upstream specification is locally acceptable but insufficiently strong for its downstream dependents. For example, as illustrated in Figure~\ref{fig:motivation}, a weak but verifiable postcondition \texttt{S1} (line 1) may provide insufficient semantic information for synthesizing \texttt{S2} (line 4), causing semantic validity issues in the latter as well. The resulting deficiency can thus propagate along the specification dependency chain and eventually manifest as verification failures of the target properties.

Confronted with such downstream failures, existing approaches typically rely on localized repair, i,e., modifying the specification at or around the observed failure location, since they are typically \textit{forward-only} and cannot revisit finalized specifications. However, the observed failure does not necessarily indicate the actual location of the underlying defect. In the example above, repeatedly repairing \texttt{S2} cannot resolve a deficiency originating from the already-finalized \texttt{S1}. The synthesis process may therefore become trapped in a failure loop: downstream failures trigger repeated local repairs while the upstream root cause remains untouched, and the verification feedback obtained at later stages cannot be utilized to revise earlier synthesis decisions.

Motivated by these observations, we aim to propose an \textit{agentic specification synthesis} paradigm, where specification synthesis is treated as an iterative decision-making process rather than a strictly forward, one-pass pipeline. Unlike conventional approaches that finalize a specification once its processing is complete, an agent can continuously reason over the evolving set of specifications and the verification feedback obtained from downstream components. When a downstream failure suggests that an upstream specification may be insufficient, the agent can trace the specification dependencies, revisit the previously generated specification, refine it, and propagate the resulting changes to its dependents. In this sense, agentic synthesis naturally tackles the aforementioned deficiency propagation problem by enabling the agent to reconsider its previous decisions in light of newly acquired evidence and iteratively improve the specification set.


\section{Methodology}
\label{sec:approach}

\begin{figure}
    \centering
    \includegraphics[width=1\linewidth]{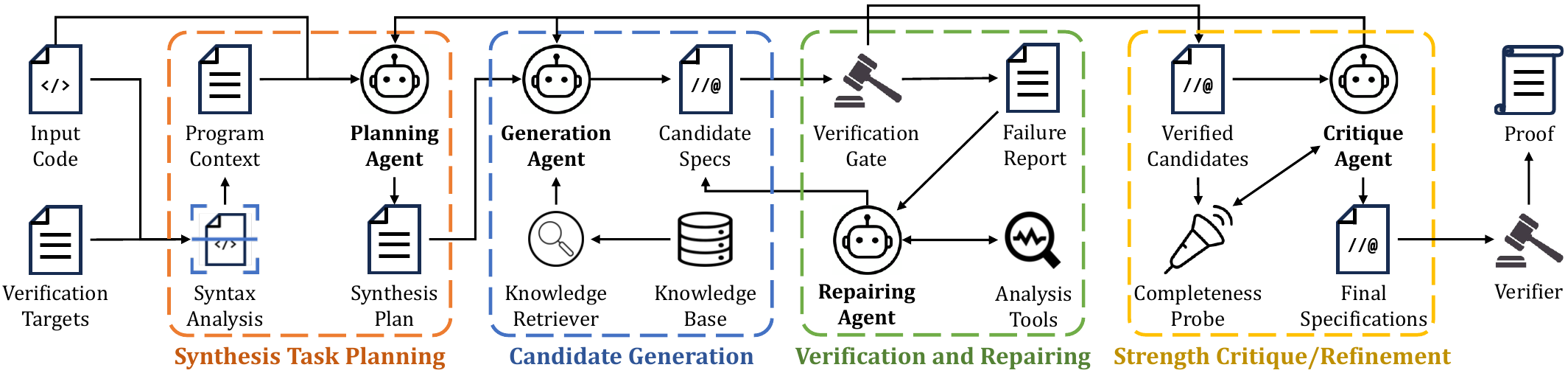}
    \vspace{-6mm}
    \caption{Overview of the proposed framework, \tool.}
    \vspace{-3mm}
    \label{fig:overview}
\end{figure}

{\noindent\bf Overview.}
\tool consists of four successive stages, as illustrated in Fig.~\ref{fig:overview}. Given a target C program, the Planning Agent first identifies required specifications, their semantics, and dependencies. The Generation Agent then constructs ACSL annotations with support from structurally relevant examples retrieved from a dedicated knowledge base. Generated candidates are checked by Frama-C/WP, and failed candidates are diagnosed and repaired by the Repairing Agent until they pass verification. Verified specifications are further assessed by the Critique Agent, which triggers refinement when insufficient properties are identified. Through collaboration among specialized agents, \tool enables flexible specification synthesis by leveraging evolving contextual information and external verification feedback.


\subsection{Specification Synthesis Task Planning}
\label{sec:planning}

Directly generating complete specifications requires an LLM to simultaneously identify relevant program properties, determine their dependencies, and express them in ACSL. Such responsibilities can become entangled for complex real-world programs, necessitating an explicit planning stage that separates specification-level reasoning from specific annotation construction. To this end, the planning Agent is designed to determine \emph{what should be specified and where to specify them} before concrete ACSL annotations are generated.

Upon receiving the target C program, \tool uses syntax analysis to summarize relevant structural information into a deterministic program context, including function signatures, caller-callee relations, and loop statement positions, before invoking the planning agent. Based on this context and direct inspection of the source code, the planning agent identifies the functions and loops that require specification instrumentation and determines the properties that should be addressed, including functional contracts and runtime safety constraints, according to the target code snippets and relevant caller/callee context. 
The output of the Planning Agent is a specification synthesis plan in natural language. The plan describes the intended specification targets, their designed semantics, interdependencies, and processing order. It therefore serves as an intermediate reasoning artifact for subsequent processes.


\subsection{Retrieval-Augmented Candidate Specification Generation}
\label{sec:generation}

\subsubsection{Candidate Specification Generation}
After the planning stage, \tool transforms the specification synthesis plan into concrete ACSL specifications by employing the \emph{Generation Agent}. Given the target program and the current planning context, the Generation Agent constructs specifications for relevant functions and loops, including preconditions (\texttt{requires}), postconditions (\texttt{ensures}), frame conditions (\texttt{assigns}), loop invariants, and predicates when necessary. The generated annotations are intended to capture the behavioral and safety properties identified during the planning stage. The Generation Agent can perform search-and-replace edits~\cite{gauthier2024aider} to instrument candidate specifications into the target code, but is constrained by the framework to specification-level modification: the agent is only allowed to add or modify ACSL annotations, without permission to modify any code implementation itself, therefore preserving the implementation semantics of the input program. The instrumented candidate specifications are then deterministically checked for syntactic validity and conformance to ACSL standards by invoking the Frama-C tool.


\subsubsection{Knowledge Base Construction and Knowledge Retrieval}
\label{sec:knowledge}
To further improve the quality of the generated candidates, \tool additionally retrieves from a meticulously constructed knowledge base a series of structurally similar code snippets with verifiable ACSL specifications instrumented, which are presented to the Generation Agent in the prompts as reference examples.

{\noindent\bf Knowledge Collection.}
To obtain specification examples grounded in reliable ACSL usage, we construct a dedicated knowledge base from the official ACSL tutorial~\cite{baudin2026acsl}. We first collect the annotated C code snippets distributed throughout the tutorial and organize them into function-level knowledge units, yielding 184 units in total. Each unit consists of a C function together with annotated specifications, preserving the original consistency between code and specification. 

\begin{figure}
    \centering
    \includegraphics[width=1\linewidth]{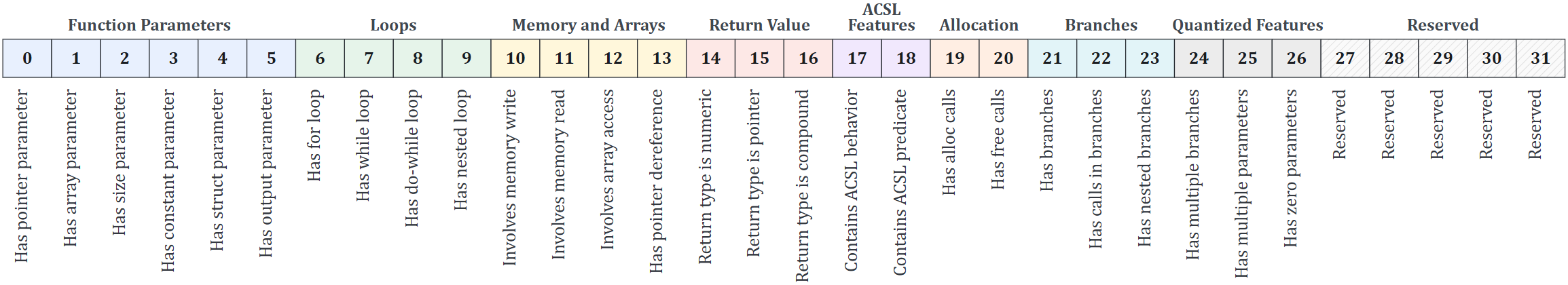}
    \vspace{-5mm}
    \caption{Illustration for the 32-bit encoding of program structural features.}
    \vspace{-3mm}
    \label{fig:encoding}
\end{figure}

{\noindent\bf Structural Fingerprint of Knowledge Units.}
For each knowledge unit, we perform syntax analysis and regex matching to extract the program's structural features. The extracted information covers a wide spectrum of aspects, including parameter types, pointer and array usage, branch and loop statements, memory accesses, function calls, return value types, and complex ACSL constructs involved. These features provide a structural description of each example and are later used to identify examples structurally compatible with the target code given. To enable efficient structural matching during retrieval, we further encode the extracted features into a compact 32-bit bitvector representation, referred to as the \emph{structural fingerprint}. Figure~\ref{fig:encoding} details the complete encoding scheme. Each bit corresponds to an aforementioned structural feature and is set when the corresponding property is present in the code snippet. The reserved bits are kept to 0 under our settings. The bitwise fingerprint enables efficient computation of the structural overlap between a target function and knowledge units. 

{\noindent\bf Knowledge Retrieval Based on Structural Similarity.}
Given a target function, \tool retrieves annotated code snippets from the knowledge base according to their structural similarity to the target. The retrieval process uses a two-stage filtering and ranking procedure, progressively narrowing the candidate space from coarse structural compatibility to fine-grained similarity. This design allows the retrieved examples to share critical program characteristics with the target while avoiding unnecessary comparisons against structurally unrelated functions.
\begin{itemize}[leftmargin=*]
    \item \tool first performs a coarse-grained structural filtering using indexed attributes in the knowledge base. Specifically, functions in candidate knowledge units are required to have compatible loop structures and return types. Knowledge units that violate this criterion are excluded from subsequent computations.
    \item The remaining candidates are further ranked according to their \textit{Jaccard Similarity} towards the target function. Denote the fingerprint of candidate function \texttt{f} as $F_\mathtt{f}$, its Jaccard Similarity towards target function \texttt{g} is calculated by
    $$
    S_J(F_\mathtt{f},F_\mathtt{g}) = \operatorname{popcount}(F_\mathtt{f}\ \&\ F_\mathtt{g}) / \operatorname{popcount}(F_\mathtt{f}\ |\ F_\mathtt{g})
    $$
    where $\operatorname{popcount}(F)=\sum_{i=0}^{31}F[i]$ denotes the number of bits set to 1 in fingerprint $F$. Eventually, up to three highest-ranked examples are selected to inject into the prompt.
\end{itemize}



\subsection{Candidate Verification and Agentic Repairing}
\label{sec:repairing}

\subsubsection{Verification Gate}
Specification synthesis is not necessarily a one-shot process, since LLM-generated specifications can be incomplete or incorrect, hindering their ability to support program verification. In \tool, each synthesized specification is subjected to a verification gate before being committed to the final set of output specifications. The verification gate encapsulates Frama-C/WP~\cite{patrick2025wp}, the verifier for ACSL, able to determine whether the candidate is syntactically valid, successfully parsed, and verifiable by the underlying verification tool. It also categorizes verification goals and outcomes reported by Frama-C/WP, including successes, syntax errors, proof failures, and timeouts, into a structured report in JSON representation, providing concrete evidence for subsequent repair. In particular, for each failure reported by the verifier, the gate maps the failure back to the source code, recovers the failure context including caller, callee, and exact source line, and reports them to the agent in detail. This turns opaque verifier-produced information into provenance-aware repair evidence, alleviating the agent's burden concerning error comprehension and root-cause reasoning.

\subsubsection{Repairing Agent}
When verification fails, a \emph{Repairing Agent} is invoked to diagnose and repair the current set of candidate specifications. Given the source code and the verification report produced by the verification gate, the agent reasons about the root cause of failure and performs corresponding repair through search-and-replace edits. Typical repairs include rectifying specifications with syntax errors or non-conformance to ACSL standards, correcting inconsistent specification semantics with the target program, and strengthening the upstream specifications with weak semantic validity. During diagnosis, the agent can also invoke Frama-C/WP and Frama-C/Eva~\cite{buhler2026eva} to obtain additional verification and value-analysis evidence, facilitating the root-cause reasoning process.
After the repairing agent finishes its job, the modified candidate is submitted to the Verification Gate again. This process iterates until the candidate passes the gate or the invocation budget for the repairing agent runs out, ensuring that all committed specifications have undergone a rigorous verification process to guarantee their correctness.

\subsection{Specification Strength Critique and Refinement}
\label{sec:critique}

\subsubsection{Quality Gate and Critique Agent}
\label{sec:quality_gate}
Verification success only establishes the consistency between the program and specifications, without guaranteeing that the specification is sufficiently \textit{strong} or \textit{complete}. \tool therefore introduces a \emph{quality gate} after successful verification to assess the semantic strength of the verified specifications. Only the verified set of specifications is subjected to this assessment, since strength is meaningful only for specifications that are correct and verifiable.

The Quality Gate invokes a \emph{Critique Agent} to perform the semantic assessment. Given the original program, synthesis plan, set of verified specifications, and relevant editing history, the agent identifies potentially weak or missing behavioral properties, including incomplete postconditions, overly restrictive preconditions, and trivial contracts. Additionally, the agent can invoke a \textit{Postcondition Completeness Probe} (detailed in Section~\ref{sec:probe}) to test whether the specification sufficiently characterizes the output of a function. Based on these observations, the agent eventually produces one of the following diagnosis results:
\begin{itemize}[leftmargin=*]
    \item \texttt{ACCEPT}: The semantic strength of current candidates is acceptable, and the framework can finish with these specifications as final output.
    \item \texttt{IMPROVE}: The specification exhibits localized strength deficiencies that do not involve interprocedural dependencies. The diagnosis report is thereby fed back to the Generation Agent for targeted regeneration.
    \item \texttt{REPLAN}: The specification exhibits structural deficiencies that involve interprocedural dependencies and may subsequently trigger cascading strength issues. Such issues are reported to the Planning Agent, which reinitiates the synthesis process anew from the planning stage.
\end{itemize}

\subsubsection{Postcondition Completeness Probe}
\label{sec:probe}

Relying solely on LLM-based reasoning to assess whether a specification sufficiently constrains program behavior can lead to inconsistent judgments. To support the Critique Agent in assessing specification strength, we introduce the \emph{Postcondition Completeness Probe} as a static analysis tool that provides stable evidence to facilitate the agent's judgment. The majority of existing approaches for specification completeness measurement are conducted through input-output pair modification~\cite{misu2026veriact} or non-equivalent program mutation~\cite{le2025can,endres2024can}, which heavily rely on the availability of representative test suites or dedicated mutation operators. Instead, we exploit a method based on the over-approximation feature of specifications~\cite{silveira2026assessing}.

The strength of a postcondition can be understood from the set of behaviors that it permits. Given a program $f$ with input space $\mathcal{I}$, let $\mathcal{B}_f=\{(i,o)\mid o=f(i), i \in \mathcal{I}\}$ denote its \emph{behavior set}, where \(i\) denotes the input program state and \(o\) denotes the observable outcome, respectively. Given a set of postconditions $S$, we similarly define its \emph{permitted behavior set} as $\mathcal{B}_S=\{(i,o)\mid S(i,o)=\mathbf{true}\}$, i.e., input-output pairs that satisfy $S$. A correct postcondition must admit all behaviors of the implementation, i.e., $\mathcal{B}_f\subseteq\mathcal{B}_S.$ Since \(S\) may permit behaviors that are not actually exhibited by \(f\), \(\mathcal{B}_S\) is an over-approximation of \(\mathcal{B}_f\).
A stronger postcondition constrains this over-approximation more tightly by excluding behaviors that are not permitted by the specification. In the ideal case, $S$ precisely characterizes the function behavior ($\mathcal{B}_f=\mathcal{B}_S$), yielding the \emph{strongest postconditions}. In this case, because a deterministic function $f$ maps each admissible input to a unique output, such a postcondition also uniquely determines the output for every admissible input.

The completeness probe checks this return-value determinacy directly through formal verification. Given a verified contract for a target function $f$, it constructs a temporary verification harness that invokes $f$ twice with identical arguments, and asserts that the two return values are equal, i.e.:
\[
m:=f(x),\qquad n:=f(x),\qquad \texttt{assert }m=n.
\]
The verifier will attempt to discharge the assertion using the target function's contract. If the verifier proves $m=n$, the contract is strong enough to uniquely determine the return value for each input, since the verifier cannot find another distinct output for the same input. Conversely, if both calls satisfy the contract's preconditions but the equality cannot be proved, the postcondition admits insufficiently constrained return behavior, providing evidence of a weak return-value specification.

Unlike prior works~\cite{silveira2026assessing,zhang2026postcondbench,misu2026veriact,le2025can,endres2024can} that primarily stop upon a mere strength assessment result, our probe is further integrated into the agent's iterative synthesis process through counterexample-guided refinement. When the aforementioned harness fails verification, the verifier-produced counterexamples and failed proof obligations are retained as part of the probe result and exposed to the Critique Agent. These artifacts provide concrete evidence about how the current specification fails to constrain the program behavior, facilitating the agent's root-cause reasoning (e.g., insufficient preconditions or under-constrained return-value characterization). The Critique Agent can thereby utilize this evidence to formulate a targeted refinement rather than merely receiving a binary judgment result. In this way, the probe serves not only as a static assessment mechanism, but also as a source of counterexample-guided evidence for specification refinement.

\begin{table}[!t]
\caption{Descriptions of all available tools in \tool.}
\label{tab:tools}
\vspace{-3mm}
\resizebox{0.92\linewidth}{!}{
\begin{tabular}{@{}c|c|c@{}}
\toprule
Tool & Description & Applicable Agent(s) \\ \midrule
\texttt{read} & Read input code or documents produced by other agents & All agents \\
\texttt{edit} & Edit candidate specifications or the document to be produced by the agent & All agents \\
\texttt{retrieve} & Retrieve similar code and specification snippets from the knowledge base & Generation \\
\texttt{check} & Check the specifications' syntactical conformance to Frama-C standards & Generation, Repairing \\
\texttt{wp} & Invoke Frama-C/WP for verification on the candidate specifications & Repairing \\
\texttt{eva} & Invoke Frama-C/Eva for auxiliary value-analysis results as reference & Repairing, Critique \\
\texttt{probe} & Invoke the Postcondition Completeness Probe & Critique \\ \bottomrule
\end{tabular}
}
\vspace{-4mm}
\end{table}

\subsection{Agent Infrastructure}
\label{sec:infra}

{\noindent\bf Agent Behavioral Paradigm.}
All agents in \tool operate under the ReAct paradigm~\cite{yao2022react}. Specifically, each agent iteratively performs the \emph{thought-action-observation} loop: in each iteration, it reasons over the currently available information, selects an appropriate action (typically a tool call), observes its outcome, and uses the obtained feedback to determine the next action. This design allows the agents to adapt their behavior (e.g., decide where to instrument specifications and what to specify) to intermediate status and results, rather than following a predefined sequential procedure (which is the common practice of existing approaches~\cite{ma2026specsyn,wang2025tale,wen2024enchanting}). The loop terminates when the agent determines that the overall objective assigned to it has been satisfied.

{\noindent\bf Tool Settings.}
All agents interact with the outer environment through tool calling. We encapsulate all necessary interactions into various tools, including file reading and writing, knowledge base query, and verifier invocation. Agents with different roles have different sets of available tools, as detailed in Table~\ref{tab:tools}. The tools are invoked during the \emph{action} step of ReAct loops through structured queries, and the corresponding outcomes are fed back to the agent as the \emph{observation} step.

{\noindent\bf Agent Orchestrator.}
The agents are coordinated by a centralized orchestrator rather than agent-authored control flow to ensure the reliability of the overall framework. The orchestrator manages the workflow progress and agent invocations through a deterministic state machine following the logic shown in Fig.~\ref{fig:overview}. Agents therefore operate within explicitly defined roles and responsibilities, without the capability to directly interfere with the overall workflow. The orchestrator is also responsible for controlling the computational budget. Each agent invocation is subject to a predefined maximum number of iterations $N_{step}$, beyond which the invocation is forcibly terminated. Also, the framework will be enforced to terminate and output current intermediate candidates as final results after the total number of agent invocations reaches a predefined parameter $N_{invoke}$.

{\noindent\bf Inter-agent Communication.}
\tool sets explicit memory boundaries between agent invocations, sharing no mutable runtime state or conversation history. Instead, agents communicate exclusively through a shared workspace. Each agent invocation can access the workspace, including current specifications, verification results, and documents produced by other agents, and can in turn update relevant workspace contents that serve as inputs to subsequent agents. Consequently, agent collaboration is achieved through successive operations on this shared intermediate context.

\section{Experimental Setup}

We aim to answer the following research questions (RQs) through experiments:
\begin{itemize}[leftmargin=*]
    \item \textbf{RQ1}: How is the general quality of the specifications generated by \tool?
    \item \textbf{RQ2}: How does each main component of \tool contributes to its performance?
    \item \textbf{RQ3}: How does \tool-generated specifications evolve with increasing agent interactions?
    \item \textbf{RQ4}: Can \tool-generated specifications facilitate real-world program verification tasks?
\end{itemize}

\subsection{Implementation}

We implement \tool as a prototype tool in Python for convenient access to LLM interfaces and agent development frameworks. Concerning the hyperparameters mentioned in Section~\ref{sec:infra}, $N_{step}$ is set to 10 for \emph{Planning} and \emph{Critique} agents, while being 30 for \emph{Generation} and \emph{Repairing} agents; the overall invocation limit $N_{invoke}$ is set to 10. Frama-C version 32.0 is adopted in the prototype and metric calculation. In the experiments, \tool and all other baselines are executed on an Ubuntu 22.04 LTS server with an Intel\textregistered\ Core\texttrademark\ Ultra 9 275HX (2.70 GHz) 24-core processor and 16GB RAM.

\subsection{Baselines}
\label{sec:baselines}

We adopt the following LLM-based approaches for ACSL specification synthesis as baselines:
\begin{itemize}[leftmargin=*]
    \item \textbf{AutoSpec}~\cite{wen2024enchanting}: One of the first specification generation approaches based on LLMs.
    \item \textbf{SpecGen}~\cite{ma2024specgen}: An approach combining LLM conversation and mutation-based repairing.
    \item \textbf{Preguss}~\cite{wang2025tale}: An approach targeting the verification of runtime error freeness properties.
    \item \textbf{SpecSyn}~\cite{ma2026specsyn}: An approach utilizing program mutation to enhance specification strength.
    \item \textbf{Codex}~\cite{openai2026codex}: A general coding agent developed by OpenAI. In the experiments, Codex is assigned permission to read/write workspace files and invoke Frama-C, while being kept from web searching. 
\end{itemize}
All baselines are backed up by two different underlying models: \texttt{deepseek-v4-pro-0813}~\cite{xu2026deepseek} and \texttt{gpt-5-2025-08-07}~\cite{openai2025gpt}. Codex is additionally equipped with \texttt{gpt-5.6-sol}~\cite{openai2026gpt}. All parameter configurations remain the same with default settings, since the adopted models do not support custom temperature settings. Note that Ma et al.~\cite{ma2026specsyn} have evaluated the performance of these baselines on GPT-5, and we directly adopt the corresponding statistics reported in their study.

\subsection{Benchmark}
\label{sec:benchmark}

{\noindent\bf RQ1\&2\&3.}
We adopt the benchmark collected by Ma et al.~\cite{ma2026specsyn} to assess the general quality of the specifications produced by baselines. The benchmark contains 50 programs collected from 14 real-world repositories, all paired with verifiable ground-truth specifications to facilitate strength assessment. Programs in this benchmark yield 456.44 LoC on average, with 7 programs exceeding 1,000 LoC. Taking into consideration that the average program size in other ACSL datasets~\cite{framacproblems,acslbyexample,cao2025informal} is typically below 50 LoC, the substantial difference in program scale makes the benchmark particularly suitable for assessing the performance of baselines on real-world programs.

{\noindent\bf RQ4.}
Evaluation on program verification tasks necessitates explicit target properties serving as verification goals. Following the settings of Ma et al.~\cite{ma2026specsyn}, we adopt two established datasets.
\begin{itemize}[leftmargin=*]
    \item Three real-world programs, collected from open-source repositories including Atomthreads~\cite{Lawson2026atomthreads}, Contiki~\cite{peyrard2018towards}, and X509-parser~\cite{ebalard2019journey}. These programs were originally collected by Wang et al.~\cite{wang2025tale} to evaluate the performance of Preguss, with target properties concerning RTE-freeness instrumented as ACSL assertions.
    \item The C category of the SV-COMP~\cite{beyer2026evaluating} benchmark. Specifically, we adopt the 213 programs in the reachability track summarized by Cao et al.~\cite{cao2025informal}, who translated the corresponding assertion targets into ACSL with other code semantics unchanged, yielding 269 targets in total.
\end{itemize}
All the aforementioned programs are directly retrieved from the artifact of Ma et al.~\cite{ma2026specsyn}.

\subsection{Metrics}
\label{sec:metrics}

Following the settings of prior works~\cite{ma2026specsyn, zhang2026postcondbench}, we adopt two distinct metrics: \emph{precision} and \emph{recall} to comprehensively evaluate different aspects of specification quality. Precision measures the \emph{correctness} of the generated specifications, calculated as the percentage of correct specifications over all generated specifications. Recall assesses the \emph{strength} of the generated specifications, measured by the percentage of ground-truth specifications \emph{covered} (i.e., there exists one generated specification that is as strong as this ground-truth) over all ground-truth specifications. We inherit the script utilized by Ma et al.~\cite{ma2026specsyn} for metric calculation, which identifies equivalence between ground-truth and generated specifications by constructing a conjunctive specification that connects the two specifications with a bidirectional implication operator and verifying its correctness with Frama-C/WP. Similar practices are widely adopted in existing works~\cite{silveira2026assessing, zhang2026postcondbench}.
\section{Experimental Results}

\subsection{RQ1: General Quality of Generated Specifications}
\label{sec:RQ1}

Table~\ref{tab:overall} demonstrates the quality (correctness measured by precision, strength measured by recall) of specifications generated by all baselines.

\begin{table}[!t]
\caption{Precision (Prec.) and recall (Rec.) of the specifications generated by each baseline, in percentages.}
\label{tab:overall}
\vspace{-3mm}
\resizebox{1\linewidth}{!}{
\begin{tabular}{ccc|cccccccccccccc|c}
\toprule
\multicolumn{3}{c|}{Repository} & X509 & Java-JML & UAV & voronoi & hypatia & sokol & cgl & par & Maratis & Punity & ccan & clib & zpl & stb & \multirow{2}{*}{Avg.} \\ \cmidrule{1-17}
\multicolumn{3}{c|}{Average LoC per File} & 95 & 207 & 1741 & 348 & 1323 & 553 & 350 & 358 & 732 & 550 & 632 & 551 & 524 & 299 &  \\ \midrule
\multicolumn{1}{c|}{\multirow{4}{*}{\tool}} & \multicolumn{1}{c|}{\multirow{2}{*}{DS-V4}} & Prec. & 97.04 & 99.20 & 89.14 & \textbf{100.00} & \textbf{100.00} & 96.88 & \textbf{100.00} & \textbf{100.00} & 80.37 & \textbf{100.00} & 97.89 & 97.82 & 97.13 & \textbf{100.00} & \textbf{96.82} \\
\multicolumn{1}{c|}{} & \multicolumn{1}{c|}{} & Rec. & 89.50 & \textbf{95.42} & \textbf{98.37} & 88.53 & 90.66 & 73.47 & 83.33 & \textbf{95.70} & \textbf{86.25} & \textbf{89.47} & 89.78 & \textbf{89.37} & \textbf{89.94} & \textbf{85.50} & \textbf{88.95} \\ \cmidrule{2-18} 
\multicolumn{1}{c|}{} & \multicolumn{1}{c|}{\multirow{2}{*}{GPT-5}} & Prec. & \textbf{100.00} & \textbf{100.00} & 76.95 & 97.26 & 100.00 & 99.30 & 100.00 & 100.00 & 81.39 & 100.00 & 99.25 & 96.73 & 97.69 & 96.88 & 96.10 \\
\multicolumn{1}{c|}{} & \multicolumn{1}{c|}{} & Rec. & 86.43 & 93.61 & 88.58 & \textbf{92.38} & \textbf{96.50} & 63.38 & 83.33 & 79.13 & 79.32 & 85.96 & 86.45 & 89.24 & 81.82 & 52.65 & 82.77 \\ \midrule
\multicolumn{1}{c|}{\multirow{4}{*}{SpecSyn}} & \multicolumn{1}{c|}{\multirow{2}{*}{DS-V4}} & Prec. & 83.81 & 91.10 & 99.59 & 96.28 & 87.54 & 60.68 & 100.00 & 90.01 & 60.58 & 100.00 & 79.17 & 76.10 & 58.78 & 93.40 & 84.07 \\
\multicolumn{1}{c|}{} & \multicolumn{1}{c|}{} & Rec. & 59.08 & 67.98 & 69.00 & 88.99 & 66.93 & 50.19 & 79.17 & 43.07 & 33.91 & 75.44 & 52.81 & 57.01 & 49.94 & 68.99 & 61.61 \\ \cmidrule{2-18} 
\multicolumn{1}{c|}{} & \multicolumn{1}{c|}{\multirow{2}{*}{GPT-5}} & Prec. & 95.66 & 97.80 & 97.57 & 99.39 & 97.24 & 93.42 & 100.00 & 98.18 & 99.13 & 98.77 & 99.24 & 90.94 & 86.88 & 99.31 & 96.68 \\
\multicolumn{1}{c|}{} & \multicolumn{1}{c|}{} & Rec. & 85.30 & 82.27 & 41.96 & 86.83 & 95.67 & \textbf{76.77} & 95.83 & 73.09 & 84.53 & 47.37 & 80.58 & 77.43 & 73.48 & 61.62 & 75.91 \\ \midrule
\multicolumn{1}{c|}{\multirow{4}{*}{AutoSpec}} & \multicolumn{1}{c|}{\multirow{2}{*}{DS-V4}} & Prec. & 83.33 & 81.96 & \textbf{100.00} & 100.00 & 100.00 & \textbf{100.00} & 100.00 & 100.00 & \textbf{100.00} & 100.00 & 90.35 & \textbf{98.61} & 74.36 & 83.34 & 93.71 \\
\multicolumn{1}{c|}{} & \multicolumn{1}{c|}{} & Rec. & 55.58 & 35.83 & 17.72 & 35.46 & 0.39 & 6.72 & 16.67 & 13.34 & 14.75 & 17.54 & 19.17 & 13.46 & 20.30 & 16.67 & 20.26 \\ \cmidrule{2-18} 
\multicolumn{1}{c|}{} & \multicolumn{1}{c|}{\multirow{2}{*}{GPT-5}} & Prec. & 33.33 & 84.72 & 100.00 & 80.00 & 95.24 & 81.25 & 42.86 & 85.40 & 16.67 & 90.00 & 66.67 & 79.39 & 81.67 & 0.00 & 66.94 \\
\multicolumn{1}{c|}{} & \multicolumn{1}{c|}{} & Rec. & 9.52 & 28.21 & 15.85 & 31.62 & 1.57 & 6.72 & 66.67 & 10.31 & 0.84 & 14.04 & 11.98 & 21.56 & 11.87 & 0.00 & 16.48 \\ \midrule
\multicolumn{1}{c|}{\multirow{4}{*}{SpecGen}} & \multicolumn{1}{c|}{\multirow{2}{*}{DS-V4}} & Prec. & 100.00 & 99.35 & 65.64 & 94.79 & 100.00 & 97.37 & 100.00 & 100.00 & 84.62 & 100.00 & 99.60 & 94.77 & 98.68 & 97.62 & 95.17 \\
\multicolumn{1}{c|}{} & \multicolumn{1}{c|}{} & Rec. & 86.62 & 93.61 & 69.23 & 50.00 & 68.09 & 40.54 & 83.33 & 93.34 & 77.00 & 89.47 & 64.09 & 74.15 & 81.20 & 71.88 & 74.47 \\ \cmidrule{2-18} 
\multicolumn{1}{c|}{} & \multicolumn{1}{c|}{\multirow{2}{*}{GPT-5}} & Prec. & 75.97 & 76.50 & 53.85 & 100.00 & 69.67 & 91.89 & 100.00 & 90.58 & 97.67 & 93.85 & 69.84 & 78.63 & 80.05 & 88.71 & 83.37 \\
\multicolumn{1}{c|}{} & \multicolumn{1}{c|}{} & Rec. & 43.46 & 68.20 & 15.38 & 77.12 & 51.97 & 81.80 & \textbf{100.00} & 79.73 & 58.06 & 68.42 & 44.19 & 50.45 & 61.24 & 43.19 & 60.23 \\ \midrule
\multicolumn{1}{c|}{\multirow{4}{*}{Preguss}} & \multicolumn{1}{c|}{\multirow{2}{*}{DS-V4}} & Prec. & 66.67 & 87.50 & 97.86 & 100.00 & 100.00 & 100.00 & 100.00 & 100.00 & 90.00 & 100.00 & 83.33 & 87.14 & \textbf{99.89} & 91.18 & 93.11 \\
\multicolumn{1}{c|}{} & \multicolumn{1}{c|}{} & Rec. & 22.93 & 29.76 & 60.14 & 34.42 & 15.18 & 12.50 & 16.67 & 17.12 & 27.26 & 24.56 & 32.84 & 28.87 & 46.28 & 26.61 & 28.22 \\ \cmidrule{2-18} 
\multicolumn{1}{c|}{} & \multicolumn{1}{c|}{\multirow{2}{*}{GPT-5}} & Prec. & 66.67 & 95.78 & 99.54 & 100.00 & 99.13 & 99.00 & 90.00 & 100.00 & 97.38 & 97.96 & 81.98 & 96.56 & 89.73 & 98.15 & 93.70 \\
\multicolumn{1}{c|}{} & \multicolumn{1}{c|}{} & Rec. & 10.27 & 29.15 & 35.43 & 36.11 & 53.15 & 13.85 & 33.33 & 21.17 & 32.76 & 24.56 & 36.42 & 38.68 & 36.44 & 28.69 & 30.72 \\ \midrule
\multicolumn{1}{c|}{\multirow{6}{*}{Codex}} & \multicolumn{1}{c|}{\multirow{2}{*}{DS-V4}} & Prec. & 95.56 & 100.00 & 69.71 & 74.00 & 100.00 & 93.89 & 100.00 & 100.00 & 86.37 & 100.00 & 99.07 & 92.78 & 84.37 & 87.93 & 91.69 \\
\multicolumn{1}{c|}{} & \multicolumn{1}{c|}{} & Rec. & 76.87 & 89.35 & 87.18 & 92.38 & 80.93 & 61.16 & 83.33 & 92.26 & 81.29 & 82.46 & \textbf{94.20} & 85.76 & 85.31 & 61.38 & 82.42 \\ \cmidrule{2-18} 
\multicolumn{1}{c|}{} & \multicolumn{1}{c|}{\multirow{2}{*}{GPT-5.6}} & Prec. & 100.00 & 100.00 & 83.63 & 95.33 & 100.00 & 98.84 & 100.00 & 99.00 & 89.62 & 100.00 & \textbf{99.49} & 97.34 & 93.25 & 89.39 & 96.14 \\
\multicolumn{1}{c|}{} & \multicolumn{1}{c|}{} & Rec. & \textbf{90.65} & 94.00 & 85.78 & 91.53 & 90.66 & 62.47 & 79.17 & 71.93 & 81.85 & 84.21 & 82.69 & 87.70 & 83.58 & 67.23 & 82.39 \\ \cmidrule{2-18} 
\multicolumn{1}{c|}{} & \multicolumn{1}{c|}{\multirow{2}{*}{GPT-5}} & Prec. & 85.61 & 98.31 & 56.30 & 93.05 & 84.28 & 95.84 & 100.00 & 99.45 & 85.20 & 86.54 & 95.54 & 90.79 & 72.20 & 87.50 & 87.90 \\
\multicolumn{1}{c|}{} & \multicolumn{1}{c|}{} & Rec. & 58.16 & 92.90 & 25.41 & 85.60 & 67.70 & 64.70 & 83.33 & 69.78 & 72.35 & 73.68 & 73.68 & 86.59 & 54.72 & 44.07 & 68.05 \\ \bottomrule
\end{tabular}
}
\vspace{-2mm}
\end{table}

{\noindent\bf Overall Quality.}
\tool achieves consistently high precision while maintaining substantially higher recall than existing approaches. Under DeepSeek-V4, \tool attains an average precision of 96.82\% and recall of 88.95\%, compared with 84.07\%/61.61\% for SpecSyn, 95.17\%/74.47\% for SpecGen, and 93.11\%/28.22\% for Preguss. Notably, Codex with \texttt{GPT-5.6-sol} achieves 96.14\% precision and 82.39\% recall, still underperforming \tool with DeepSeek-V4. The same trend holds for GPT-5: \tool achieves 96.10\% precision and 82.77\% recall, surpassing all baselines. These results indicate that \tool can generate specifications that are not only predominantly correct but also substantially stronger than those produced by most existing approaches. 

\begin{figure}[!t]
    \centering
    \includegraphics[width=1\linewidth]{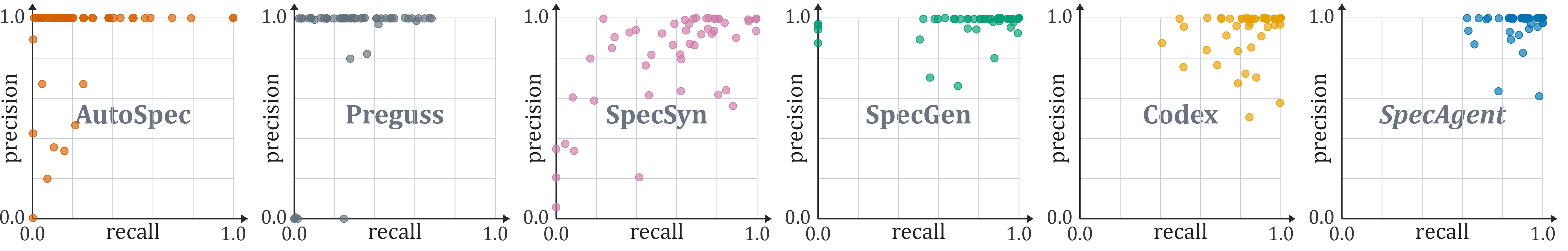}
    \vspace{-7mm}
    \caption{Per-program specification quality of each baseline (using DeepSeek-V4) shown by scatter points. The coordinates of each point represent the precision and recall achieved on one specific program.}
    \vspace{-1mm}
    \label{fig:scatter}
\end{figure}

{\noindent\bf Trade-off between Correctness and Strength.}
The results further reveal a clear trade-off faced by existing methods. AutoSpec and Preguss, for example, often achieve high precision on individual repositories, but their recalls remain low, with average recall as low as 20.26\% and 28.22\% under DeepSeek-V4, respectively. This suggests that their high precision is frequently accompanied by limited specification coverage. In contrast, \tool maintains precision above 96\% while reaching nearly 89\% recall with DeepSeek-V4. 
This trend is even more clearly illustrated in Fig.~\ref{fig:scatter}. Across all baselines, the majority of data points lie in the upper region of the grid, indicating that existing approaches generally achieve high precision. However, the points of Preguss and AutoSpec are predominantly concentrated toward the upper-left corner, whereas those of \tool are substantially more clustered around the upper-right corner, highlighting the pronounced gap in recall across different approaches.
This result is particularly notable because attempting to enhance specification strength introduces additional risk of incorrect or unverifiable assertions. The ability of \tool to improve recall without a corresponding substantial loss in precision suggests that its agentic synthesis process can exploit verification feedback to effectively enrich specification semantics while preserving correctness.

{\noindent\bf Robustness across repositories.}
\tool also demonstrates relatively stable performance across projects with substantially different program sizes. The benchmark ranges from relatively small programs such as those from X509-parser (95 LoC per file) to substantially larger ones such as UAV (1,741 LoC) and hypatia (1,323 LoC). Despite this variation, \tool maintains precision close to or above 90\% on most repositories and achieves high recall on both small and large programs. For example, on UAV, the largest repository in the benchmark, \tool obtains 89.14\% precision and 98.37\% recall with DeepSeek-V4, and 76.95\% precision and 88.58\% recall with GPT-5. This suggests that the approach is not limited to short or structurally simple programs and can retain substantial specification strength as program complexity increases.

{\noindent\bf Effect of the Underlying LLM.}
Changing the underlying LLM affects the absolute performance of all approaches, but does not alter the relative advantage of \tool. Replacing DeepSeek-V4 with GPT-5 slightly decreases average recall of \tool from 88.95\% to 82.77\%, while precision remains comparably high (96.82\% vs. 96.10\%). In contrast, several baselines exhibit much larger degradations in recall when switching models. e.g., SpecGen from 74.47\% to 60.23\%. It is noteworthy that recall of Codex drops from 82.42\% to 68.05\% (-16.37\%), whereas \tool's recall decreases by only 6.18\%, despite both Codex and \tool adopting agentic workflows. This comparison suggests that \tool's explicit verification-guided state transitions and iterative refinement make specification synthesis less sensitive to the underlying LLM's agentic capabilities. 

\vspace{-1mm}
\begin{tcolorbox}[left=1mm,right=1mm]
    \vspace{-1mm}
    \textbf{Answer to RQ1:} \tool consistently generates specifications with high correctness and strength across repositories, achieving an average precision of 96.82\% and recall of 88.95\% with DeepSeek-V4, and 96.10\% and 82.77\% with GPT-5. Compared with existing approaches, \tool better balances specification correctness and strength, avoiding the weak specifications often produced by methods that prioritize precision. Moreover, \tool exhibits relatively stable performance across different underlying LLMs, suggesting that its verification-guided agentic workflow reduces the dependence of specification quality on the underlying model.
\vspace{-1mm}
\end{tcolorbox}
\vspace{-1mm}

\subsection{RQ2: Ablation Study on Main Components}

We construct 4 variants of \tool by progressively augmenting a basic candidate generation process (Section~\ref{sec:generation}) with additional agentic components. Specifically, G-only represents the basic generation process, followed by P+G that augments it with the Planning stage (Section~\ref{sec:planning}), P+G+R further with Verification and Agentic Repairing (Section~\ref{sec:repairing}), and finally the complete \tool configuration with Strength Critique and Refinement (Section~\ref{sec:critique}).

Table~\ref{tab:ablation} reports the ablation results on DeepSeek-V4 and GPT-5. Overall, the results show that the individual components provide complementary benefits, while their combination is essential for achieving consistently strong performance. Using Generation alone (G only) already achieves relatively high precision, with average precision of 89.78\% and 88.66\% for DeepSeek-V4 and GPT-5, respectively, but its recall remains limited to 80.93\% and 72.14\%. Adding Planning (P+G) does not yield a consistent improvement: although recall increases slightly for DeepSeek-V4 to 83.99\%, it slightly decreases for GPT-5 to 71.51\%, suggesting that planning alone cannot reliably address synthesis failures. In contrast, incorporating Repairing (P+G+R) substantially improves precision, reaching 94.79\% and 94.34\% on average for DeepSeek-V4 and GPT-5, respectively, while also recovering recall to 82.08\% and 76.21\%. Finally, the complete \tool configuration achieves the strongest overall performance, with average precision/recall of 96.82\%/88.95\% for DeepSeek-V4 and 96.10\%/82.77\% for GPT-5. Notably, the gain in recall over Generation-only is particularly pronounced for \tool, reaching +8.02\% and +10.63\% for DeepSeek-V4 and GPT-5, respectively. These results indicate that the key advantage of \tool does not stem from any single component, but from their interaction: Planning provides dependency-aware guidance, Repairing resolves invalid or unverifiable specifications, and the complete workflow enables these decisions to be iteratively informed by verification feedback.

\begin{table}[!t]
\caption{Performance of all ablation variants of \tool. All statistics are reported in percentages. P: \emph{synthesis task Planning}, G: \emph{candidate Generation}, R: \emph{agentic Repairing}, C: \emph{strength Critique and refinement}.}
\label{tab:ablation}
\vspace{-3mm}
\resizebox{1\linewidth}{!}{
\begin{tabular}{@{}c|c|c|cccccccccccccc|c@{}}
\toprule
Model & Variant & Metric & X509 & Java-JML & UAV & voronoi & hypatia & sokol & cgl & par & Maratis & Punity & ccan & clib & zpl & stb & Avg. \\ \midrule
\multirow{8}{*}{DS-V4} & \multirow{2}{*}{G only} & Prec. & 95.24 & 91.25 & 66.55 & 64.70 & 86.40 & \textbf{98.15} & 100.00 & 98.99 & \textbf{91.67} & 92.68 & 99.48 & 74.80 & 97.06 & 100.00 & 89.78 \\
 &  & Rec. & 83.21 & 94.06 & 82.28 & 88.14 & 90.66 & 68.65 & 83.33 & 94.16 & 47.46 & 78.95 & 81.61 & 70.86 & 84.10 & 85.50 & 80.93 \\ \cmidrule(l){2-18} 
 & \multirow{2}{*}{P+G} & Prec. & 94.67 & 90.38 & 66.18 & 66.60 & 90.47 & 97.29 & 96.76 & 95.40 & 86.37 & 95.78 & 96.25 & \textbf{99.04} & 91.83 & 95.65 & 90.19 \\
 &  & Rec. & 86.43 & 95.92 & 79.72 & \textbf{91.53} & 90.66 & 52.14 & 83.33 & \textbf{98.98} & 86.25 & 89.47 & 81.25 & 82.47 & 88.94 & 68.83 & 83.99 \\ \cmidrule(l){2-18} 
 & \multirow{2}{*}{P+G+R} & Prec. & \textbf{100.00} & \textbf{100.00} & 73.97 & 94.16 & 100.00 & 95.00 & 100.00 & 100.00 & 83.85 & 100.00 & \textbf{99.59} & 87.30 & 93.18 & 100.00 & 94.79 \\
 &  & Rec. & 84.10 & 84.33 & 88.67 & 77.24 & 87.93 & 64.25 & 77.50 & 89.41 & 87.33 & 83.21 & 85.42 & 80.13 & 80.04 & 79.52 & 82.08 \\ \cmidrule(l){2-18} 
 & \multirow{2}{*}{Full} & Prec. & 97.04 & 99.20 & \textbf{89.14} & \textbf{100.00} & \textbf{100.00} & 96.88 & \textbf{100.00} & \textbf{100.00} & 80.37 & \textbf{100.00} & 97.89 & 97.82 & \textbf{97.13} & \textbf{100.00} & \textbf{96.82} \\
 &  & Rec. & \textbf{89.50} & \textbf{95.42} & \textbf{98.37} & 88.53 & \textbf{90.66} & \textbf{73.47} & \textbf{83.33} & 95.70 & \textbf{86.25} & \textbf{89.47} & \textbf{89.78} & \textbf{89.37} & \textbf{89.94} & \textbf{85.50} & \textbf{88.95} \\ \midrule
\multirow{8}{*}{GPT-5} & \multirow{2}{*}{G only} & Prec. & 91.67 & 97.53 & 72.46 & 70.33 & 86.44 & 86.26 & 100.00 & 96.07 & \textbf{81.49} & 98.48 & 93.82 & 95.54 & 88.41 & 82.69 & 88.66 \\
 &  & Rec. & 64.03 & 86.82 & 71.33 & 88.14 & 67.70 & 56.80 & 83.33 & 74.04 & 71.51 & 78.95 & 69.74 & 74.13 & 80.48 & 43.03 & 72.14 \\ \cmidrule(l){2-18} 
 & \multirow{2}{*}{P+G} & Prec. & 89.93 & 98.69 & 73.98 & 60.45 & 61.26 & 89.91 & 100.00 & 100.00 & 87.73 & 94.59 & 94.17 & 97.09 & 90.91 & 86.45 & 87.51 \\
 &  & Rec. & 63.65 & 87.88 & 51.98 & 88.14 & 59.14 & 58.52 & 83.33 & 73.56 & 72.23 & 82.46 & 66.44 & 81.22 & 77.11 & 55.53 & 71.51 \\ \cmidrule(l){2-18} 
 & \multirow{2}{*}{P+G+R} & Prec. & 100.00 & 100.00 & 68.82 & 71.67 & 100.00 & 96.38 & 100.00 & 100.00 & 85.50 & 100.00 & \textbf{100.00} & \textbf{99.76} & \textbf{98.69} & \textbf{100.00} & 94.34 \\
 &  & Rec. & 75.94 & 88.86 & 82.28 & 85.14 & 87.94 & 56.80 & 83.33 & 71.70 & 75.76 & 80.70 & 69.22 & 80.69 & 69.19 & \textbf{59.38} & 76.21 \\ \cmidrule(l){2-18} 
 & \multirow{2}{*}{Full} & Prec. & \textbf{100.00} & \textbf{100.00} & \textbf{76.95} & \textbf{97.26} & \textbf{100.00} & \textbf{99.30} & \textbf{100.00} & \textbf{100.00} & 81.39 & \textbf{100.00} & 99.25 & 96.73 & 97.69 & 96.88 & \textbf{96.10} \\
 &  & Rec. & \textbf{86.43} & \textbf{93.61} & \textbf{88.58} & \textbf{92.38} & \textbf{96.50} & \textbf{63.38} & \textbf{83.33} & \textbf{79.13} & \textbf{79.32} & \textbf{85.96} & \textbf{86.45} & \textbf{89.24} & \textbf{81.82} & 52.65 & \textbf{82.77} \\ \bottomrule
\end{tabular}
}
\vspace{-4mm}
\end{table}

\vspace{-1mm}
\begin{tcolorbox}[left=1mm,right=1mm]
    \vspace{-1mm}
    \textbf{Answer to RQ2:} The ablation study shows that the components of \tool provide complementary benefits, with performance improving as the synthesis process is progressively augmented with planning, repairing, and critique. In particular, repairing substantially improves precision, while the complete workflow further improves recall by leveraging the feedback from strength Critique. Overall, the results demonstrate the importance of integrating these components into an agentic synthesis process rather than relying on a single agent.
\vspace{-1mm}
\end{tcolorbox}
\vspace{-1mm}

\subsection{RQ3: Specification Quality Evolution over Agent Invocations}
\label{sec:evolution}

To investigate how specification quality evolves as agent interactions accumulate, we analyze all specification evolution trajectories at increasing numbers of agent invocations. Specifically, we vary the number of agent invocations from 2 to 10 (1 excluded since the generation stage has not yet begun) and track four metrics at each invocation count: the fraction of trajectories that have passed the verification gate, the average number of generated specifications, precision, and recall. The first metric captures how quickly trajectories reach a verification-accepted state, while the latter three characterize the quantity and quality of the resulting specifications. This analysis allows us to examine how additional agent interactions contribute to specification synthesis and whether their benefits diminish as the invocation count increases.

\begin{figure}[!t]
    \centering
    \includegraphics[width=1\linewidth]{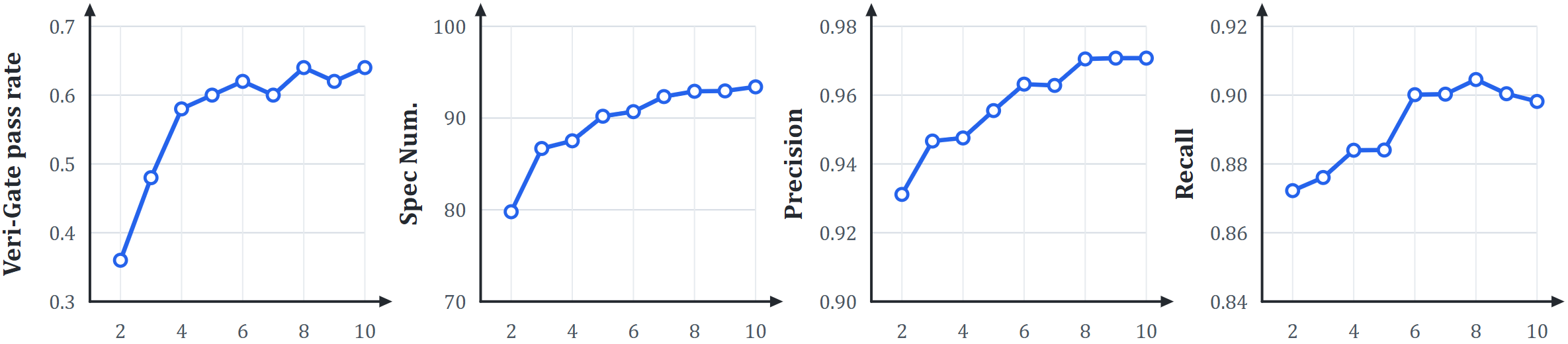}
    \vspace{-7mm}
    \caption{Evolution of \tool-generated specification quality as agent invocations increase from 2 to 10.}
    \vspace{-3mm}
    \label{fig:evolution}
\end{figure}

Fig.~\ref{fig:evolution} demonstrates the trend of candidate specification quality evolution as the number of agent invocations increases. As the number of agent invocations increases, candidate specification quality generally improves, but the gains gradually diminish. The most substantial improvement occurs in the early invocations: from 2 to 6 invocations, the mean precision increases from 93.11\% to 96.32\%, while the mean recall rises from 87.22\% to 90.01\%. Meanwhile, the average number of generated specifications increases from 79.79 to 90.70, indicating that additional agent interactions progressively expand the set of candidate specifications while maintaining or improving their quality. Most notably, the fraction of trajectories passing the verification gate also increases from 36\% at 2 invocations to 62\% at 6 invocations, suggesting that 2--6 is the golden interval where the majority of trajectories reach verifiable states. Beyond 6--8 invocations, however, the improvements become marginal: precision remains around 97\%, recall fluctuates around 90\%, and the verification-gate pass rate oscillates between 60\% and 64\%. These results suggest that agentic specification synthesis benefits substantially from a limited number of iterative interactions, after which additional invocations yield diminishing returns rather than continuously improving specification quality. This diminishing-return trend also justifies our choice of 10 as the maximum agent invocation budget, as further interactions are unlikely to yield substantial improvements in specification quality.

\vspace{-1mm}
\begin{tcolorbox}[left=1mm,right=1mm]
    \vspace{-1mm}
    \textbf{Answer to RQ3:} Agentic specification synthesis exhibits clear diminishing returns with increasing invocation budgets. Most improvements occur within the first 6--8 invocations, after which these metrics largely stabilize around 97\% precision, 90\% recall, and 60--64\% verification-gate pass rate. This indicates that a moderate invocation budget is sufficient for the agent to capture most of the attainable specification quality.
\vspace{-1mm}
\end{tcolorbox}
\vspace{-1mm}

\subsection{RQ4: Performance on Real-world Program Verification Tasks}
\label{sec:RQ4}

To assess the practical effectiveness of the evaluated baselines on real-world software verification tasks, we evaluate all approaches on two types of real-world verification tasks. Table~\ref{tab:verification} demonstrates the number of verification targets successfully discharged by all baseline methods. Overall, \tool successfully supports the verification of 604 out of 811 targets (74.5\%), compared with 523 for Codex, 425 for Preguss, and 468 for SpecSyn. This result shows that \tool can effectively translate program context into specifications strong enough to discharge a large fraction of verification targets. The following analysis further examines this performance across the two types of real-world verification tasks.

{\noindent\bf X509-parser, Contiki, and Atomthreads.}
The programs are extracted from real-world repositories, with ACSL assertions on runtime error (RTE) freeness instrumented as verification targets. For these programs, \tool consistently enables the successful verification of a large fraction of the target properties. Notably, on X509-parser, \tool successfully discharges all 140 targets, compared with 125 and 126 for Preguss and SpecSyn, respectively. On Contiki, \tool verifies 137 out of 163 targets, again exceeding Preguss (103) and SpecSyn (105). On the more challenging Atomthreads project, \tool successfully verifies 109 of 239 targets, comparable to Preguss (109) but above SpecSyn and Codex (99). As described in Section~\ref{sec:benchmark}, the programs were originally intended to evaluate Preguss, an approach focusing on excluding potential RTEs. The relatively stable performance of Preguss across these programs suggests its effectiveness on verification targets concerning RTE-freeness. Codex provides a relatively competitive baseline, but its performance remains below \tool and Preguss, indicating that a general-purpose agent alone may not consistently produce specifications sufficient for verification. It is worth noting that SpecGen verifies none of the targets on X509 and only a small fraction on Contiki and Atomthreads. This is primarily due to its generation strategy, which requires the LLM to reproduce the input program while inserting ACSL annotations into the output, introducing the risk of unintended source-code modifications. In our experiments, SpecGen modified the source code of both Contiki and Atomthreads during generation, rendering the corresponding verification targets invalid. 

{\noindent\bf SV-COMP.}
The programs retrieved from the SV-COMP benchmark are mainly toy examples (average LoC being 27.79), on which the verification tasks primarily focus on the synthesis of loop invariants. Among 269 targets, \tool successfully verifies 218, substantially outperforming SpecSyn (138) and Preguss (88), while remaining competitive with Codex (206), indicating that \tool possesses a strong capability to reason about iterative program behavior and synthesize inductive invariants that capture non-trivial loop semantics. In contrast, Preguss and SpecSyn, which perform relatively competitively on some real-world repositories, show substantially lower coverage on SV-COMP, suggesting that their strengths on real-world RTE-oriented properties may not transfer equally well to loop-invariant synthesis. Codex, in contrast, achieves 206 successful verifications and remains close to \tool, indicating that its general-purpose code reasoning capability is particularly effective on the relatively small SV-COMP programs. 

\begin{table}[!t]
\caption{Number of Verified Targets yielded by each baseline.} 
\label{tab:verification}
\vspace{-3mm}
\resizebox{0.83\linewidth}{!}{
\begin{tabular}{@{}c|c|c|cccccc@{}}
\toprule
Repository & Avg. LoC & Target Num. & \tool & Preguss & SpecSyn & SpecGen & AutoSpec & Codex \\ \midrule
X509-parser & 1199 & 140 & \textbf{140} & 125 & 126 & 41 & 81 & 113 \\
Contiki & 544 & 163 & \textbf{137} & 103 & 105 & 43 & 89 & 105 \\
Atomthreads & 1451 & 239 & \textbf{109} & 109 & 99 & 27 & 7 & 99 \\ \midrule
SV-COMP & 27.79 & 269 & \textbf{218} & 88 & 138 & 90 & 28 & 206 \\ \midrule
Total & - & 811 & \textbf{604} & 425 & 468 & 201 & 205 & 523 \\ \bottomrule
\end{tabular}
}
\vspace{-4mm}
\end{table}

\vspace{-1mm}
\begin{tcolorbox}[left=1mm,right=1mm]
    \vspace{-1mm}
    \textbf{Answer to RQ4.} \tool achieves the highest overall verification coverage, successfully discharging 604 out of the 811 verification targets and consistently performing strongly across real-world projects and SV-COMP tasks. The results demonstrate that \tool can synthesize sufficiently strong specifications for both large-scale real-world code and loop-intensive verification tasks, indicating its ability to adapt the synthesis procedure of different types of specifications to diverse verification requirements.
\vspace{-1mm}
\end{tcolorbox}
\vspace{-1mm}

\section{Discussion}


\subsection{Cost Analysis}

We further analyze the computational and monetary overhead of the evaluated approaches in terms of wall-clock time, Frama-C invocations, LLM calls, and API cost. As shown in Table~\ref{tab:cost}, \tool requires 1,684.49 seconds and 25.58 Frama-C calls per program, far more efficient than verification-intensive approaches (Preguss and SpecSyn). Although the wall-time overhead over lightweight approaches (AutoSpec and SpecGen) is non-negligible, it remains substantially lower than the human effort required for real-world verification projects (e.g., developing ACSL specifications and completing the verification of X509-parser required five calendar months of manual effort~\cite{ebalard2019journey}). Moreover, \tool requires 99.40 LLM calls, slightly fewer than Preguss (107.30), with monetary cost being \$0.6228, lower than Preguss (\$0.7210) and Codex (\$0.7100) while comparable to SpecSyn (\$0.6243). Among the baselines, AutoSpec achieves the lowest overhead across all metrics. However, this efficiency is accompanied by substantially weaker effectiveness in specification generation and program verification. SpecGen also maintains low computational overhead but exhibits high instability on verification tasks (Section~\ref{sec:RQ4}). In contrast, Preguss and SpecSyn incur considerably higher verification overhead, with 351.22 and 369.33 Frama-C calls, respectively, resulting in substantially longer wall time as well. Overall, \tool achieves a practical balance between verification effectiveness and resource consumption, introducing moderate overhead to enable dependency-aware, verification-guided, and iterative specification synthesis.

\begin{table}[!t]
\caption{Average execution overhead per program for each baseline on DeepSeek-V4. We calculate the monetary costs from the figures reported by the official DeepSeek API platform~\cite{deepseek2026platform}.}
\label{tab:cost}
\vspace{-3mm}
\resizebox{0.67\linewidth}{!}{
\begin{tabular}{@{}ccccccc@{}}
\toprule
Metrics & Preguss & SpecGen & SpecSyn & AutoSpec & Codex & \tool \\ \midrule
Wall Time (s) & 7750.77 & 898.88 & 10305.96 & 582.46 & 1717.72 & 1684.49 \\
Frama-C Calls & 351.22 & 5.04 & 369.33 & 3.76 & 2.28 & 25.58 \\
LLM Calls & 107.30 & 4.46 & 74.82 & 3.43 & 120.02 & 99.40 \\
Cost (USD) & 0.7210 & 0.3852 & 0.6243 & 0.0341 & 0.7100 & 0.6228 \\ \bottomrule
\end{tabular}
}
\vspace{-4mm}
\end{table}


\subsection{Threats to Validity}

{\noindent\bf Internal Threats.}
First, LLM-based agents may produce different outputs across runs due to the stochastic nature of generation, introducing potential variance into specification qualities. We mitigate this threat by fixing the model configurations across experiments and reporting aggregate results over a series of evaluated tasks.
Second, our ablation study incrementally adds major components to the basic candidate-generation workflow, but adding a component may also change the subsequent execution trajectory and the number of agent interactions, making it difficult to completely isolate its individual contribution. We therefore interpret the ablation results as measuring the contribution of each component within the overall workflow rather than as strictly independent causal effects.
Finally, the effectiveness of \tool may be sensitive to agent budget configurations, e.g., the maximum number of agent invocations. Concerning this threat, our additional study on specification quality evolution shows that specification quality improves noticeably within the first 6–8 agent invocations, after which it tends to plateau, indicating that insufficient budgets can noticeably degrade specification quality, while larger budgets yield diminishing returns. The analysis provides empirical insights into the sensitivity of \tool to agent budget and how to configure the budget in practical scenarios.

{\noindent\bf External Threats.}
The main external threat lies in the benchmark adopted for experiments. Although the benchmark programs come from real-world open-source repositories, the number of programs in the adopted benchmark is still limited. We mitigate this threat by evaluating the baselines across multiple repositories and complementing the experiments with real-world verification tasks. Further evaluation on larger and more diverse software systems is left for future work.
Another threat concerns the language scope, as our evaluation focuses on C programs specified in ACSL and verified with the Frama-C/WP ecosystem. The effectiveness of the proposed agentic workflow may depend on the characteristics of this programming language and formal verification ecosystem, and therefore our findings may not directly generalize to other specification languages and programming languages.
Lastly, our evaluation relies on Frama-C/WP to assess specification correctness and strength. The verification tool is inherently incomplete, and a valid specification may fail to be proved due to limitations of the underlying solver, leading to potential false negatives in our evaluation. Nevertheless, we employ the same Frama-C/WP version, verification configuration, and underlying solver environment across all experiments and baselines, ensuring the consistency and fairness of the comparisons.
\section{Related Works}

{\noindent\bf LLM-based Agent for Software Engineering.}
Recent advances in LLMs have shifted software engineering automation from mere LLM queries toward autonomous agents that iteratively interact with software environments. Early efforts such as MetaGPT~\cite{hong2024metagpt} and ChatDev~\cite{qian2024chatdev} focus on code generation by organizing software development as collaboration among role-specialized agents through predefined workflows for design, coding, and testing. As agentic capabilities evolve, recent works extend to repository-level tasks involving existing software. SWE-agent~\cite{yang2024swe} addresses real-world GitHub issues by enabling agents to navigate repositories, edit files, and execute tests through a specialized interface. AutoCodeRover~\cite{zhang2024autocoderover} incorporates program structures and iterative code search for issue localization and program repair. RepairAgent~\cite{bouzenia2025repairagent} enables autonomous defect repair by allowing LLMs to gather information, generate patches, invoke validation tools, and adapt actions based on feedback. OpenHands~\cite{wang2025openhands} further explores general-purpose software engineering agents that interact with code, command-line tools, and development environments. Despite this progress, agentic approaches for formal methods remain relatively nascent. Tu et al.~\cite{tu2026agentic} introduce AutoRocq, an LLM agent that performs program verification in the Rocq prover by iteratively refining proofs based on prover feedback. Sun et al.~\cite{sun2026agentic} employ LLM agents to orchestrate model checking tasks by selecting strategies, analyzing counterexamples, and refining decisions based on checker feedback. Compared to these works, \tool treats formal specification synthesis as the primary task, perform agentic generation, repair, and refinement on formal specifications.

{\noindent\bf LLM-based Program Specification Synthesis.}
The emergence of LLMs has enabled specification synthesis beyond conventional rule-based and analysis-based approaches. AutoSpec~\cite{wen2024enchanting} pioneers LLM-based specification generation by leveraging static analysis to expose program hierarchies and guide synthesis. SpecGen~\cite{ma2024specgen} combines conversational generation with heuristic mutation operators to repair generated specifications, while Preguss~\cite{wang2025tale} employs abstract interpretation to detect potential runtime errors and guide generation. SpecSyn~\cite{ma2026specsyn} further utilizes non-equivalent program mutations to assess and refine the semantic strength of generated specifications. Pei et al.~\cite{pei2023can} and Cao et al.~\cite{cao2025informal} improve specification generation by fine-tuning LLMs on specification-related corpora. Compared to these works, \tool reformulates specification synthesis as an agentic decision-making process, where specialized agents dynamically determine generation, repair, and refinement actions based on evolving context. Other works~\cite{chen2024automated,yang2025autoverus,zhong2025rag,jiang2026starverus} target different specification languages, such as Verus for Rust. Due to substantial differences in programming languages and formal verification ecosystems, we do not directly compare \tool with these approaches. Among them, StarVerus~\cite{jiang2026starverus}, a concurrent work, explores autonomous agents for synthesizing and repairing formal specifications using verifier feedback. However, its primary objective is to obtain specifications that pass verification, without explicitly critiquing their strength. In contrast, \tool treats specification correctness as a prerequisite rather than the endpoint, explicitly identifying semantic deficiencies in the specifications and refining them when necessary.

\section{Conclusion}

We present \tool, an agentic framework for automated ACSL specification synthesis that reformulates specification generation as an iterative decision-making process. By integrating planning, generation, verification-guided repair, and critique-guided refinement, \tool identifies and revisits specification dependencies rather than following a fixed forward-only workflow. Extensive evaluation on real-world specification generation and verification tasks demonstrates that \tool improves specification completeness while maintaining high precision, effectively supporting practical software verification. These results highlight the potential of agentic approaches for reliable and scalable formal specification synthesis.


\bibliography{ref}
\bibliographystyle{plainnat}

\end{document}